\documentclass[preprint,nofootinbib,aps,superscriptaddress,eqsecnum]{revtex4-1} 
 \pdfoutput=1
 \usepackage{graphicx}
 \usepackage{amsmath}
\usepackage{amsfonts}
\usepackage{amssymb}
\usepackage{hyperref}
\usepackage{gensymb}

\usepackage{caption}
\usepackage{subcaption}
\def\bea{\begin{eqnarray}}
\def\eea{\end{eqnarray}}
 \def\be{\begin{equation}}
\def\ee{\end{equation}}

\begin{document}

\title{ Constraints on long range force from perihelion precession of planets in a gauged $L_e-L_{\mu,\tau}$ scenario }

  \author{Tanmay Kumar Poddar}
\email[Email Address: ]{tanmay@prl.res.in}
\affiliation{Theoretical Physics Division, 
Physical Research Laboratory, Ahmedabad - 380009, India}
\affiliation{Discipline of Physics, Indian Institute of Technology, Gandhinagar - 382355, India}

\author{Subhendra Mohanty}
\email[Email Address: ]{mohanty@prl.res.in}
\affiliation{Theoretical Physics Division, 
Physical Research Laboratory, Ahmedabad - 380009, India}

\author{Soumya Jana }
\email[Email Address: ]{Soumya.Jana@etu.unige.ch}
\affiliation{\it D\'epartement de Physique Th\'eorique, Universit\'e de Gen\`eve, 24 quai Ernest Ansermet, 1211Gen\`eve 4, Switzerland}
\affiliation{ Department of Physics, Sitananda College, Nandigram, 721631, India} 

\begin{abstract}
The standard model particles can be gauged in an anomaly free way by three possible gauge symmetries namely ${L_e-L_\mu}$, ${L_e-L_\tau}$, and ${L_\mu-L_\tau}$. Of these, ${L_e-L_\mu}$ and ${L_e-L_\tau}$ forces can mediate between the Sun and the planets and change the perihelion precession of planetary orbits. It is well known that a deviation from the $1/r^2$ Newtonian force can give rise to a perihelion advancement in the planetary orbit, for instance, as in the well known case of Einstein's gravity which was tested from the observation of the perihelion advancement of the Mercury. We consider the long range Yukawa potential which arises between the Sun and the planets if the mass of the gauge boson is $M_{Z^{\prime}}\leq \mathcal{O}(10^{-19})\rm {eV}$. We derive the formula of perihelion advancement for Yukawa type fifth force due to the mediation of such $U(1)_{L_e-L_{\mu,\tau}}$ gauge bosons. The perihelion advancement for Yukawa potential is proportional to the square of the semi major axis of the orbit for small $M_{Z^{\prime}}$, unlike GR, where it is largest for the nearest planet. However for higher values of $M_{Z^{\prime}}$, an exponential suppression of the perihelion advancement occurs. We take the observational limits for all planets for which the perihelion advancement is measured and we obtain the upper bound on the gauge boson coupling $g$ for all the planets. The Mars gives the stronger bound on $g$ for the mass range $\leq 10^{-19}\rm{eV}$ and we obtain the exclusion plot. This mass range of gauge boson can be a possible candidate of fuzzy dark matter whose effect can therefore be observed in the precession measurement of the planetary orbits. 
\end{abstract}

\maketitle

\section{Introduction}
It is well known that a deviation from the inverse square law force between the Sun and the planets results in the perihelion precession of the planetary orbits around the Sun. One of the most prominent example is the case of the Einstein's general relativity (GR) which predicts a deviation from Newtonian $1/r^2$ gravity. In fact, one of the famous classical tests of GR was to explain the perihelion advancement of the Mercury. There was a mismatch of about $43$ arc seconds per century from the observation \cite{shapiro1990} which could not be explained from Newtonian mechanics by considering all non-relativistic effects such as perturbations from the other Solar System bodies, oblateness of the Sun, etc. GR explains the discrepancy with a prediction of contribution of $42.9799''/$Julian century \cite{park}. However there is an uncertainty in the GR prediction which is about $10^{-3}$ arc seconds per century \cite{park,genova,iorio,Sun} for the Mercury orbit. The current most accurate detection of perihelion precession of Mercury is done by MESSENGER mission \cite{genova}. In the near future, more accurate results will come from BepiColombo mission \cite{will}. Other planets also experience such perihelion shift, although the shifts are small since they are at larger distance from the Sun \cite{biswas,liorio}. 


 The uncertainty in GR prediction opens up the possibility to explore the existence of Yukawa type potential between the Sun and the planets leading to the fifth force which is a deviation from the inverse-square law.  Massless or ultralight scalar, pseudoscalar or vector particles can mediate such fifth force between the Sun and the planets. Many recent papers constrain the fifth force originated from either scalar-tensor theories of gravity \cite{liu,janafr,alexander} or the dark matter components \cite{croon,laha,alexander}. Fifth forces due to ultra light axions was studied in \cite{jana}. Ultra light scalar particles can also be probed from the coupling of electron in long range force effects in torsion balance experiment \cite{garv}. They can also be probed from superradiance phenomena \cite{teo,denton}. The unparticle long range force from perihelion precession of Mercury was studied in \cite{suratna}. Perihelion precession of planets can also constrain the fifth force of dark matter \cite{Sun}. In this paper, we consider the Yukawa type potential which arises in a gauged $L_e-L_{\mu,\tau}$ scenario and we calculate the perihelion shift of planets (Mercury, Venus, Earth, Mars, Jupiter, and Saturn) due to coupling of the ultralight vector gauge bosons with the electron current of the macroscopic objects along with the GR effect.
 
\begin{figure}[!htbp]
\centering
\includegraphics[width=4.0in]{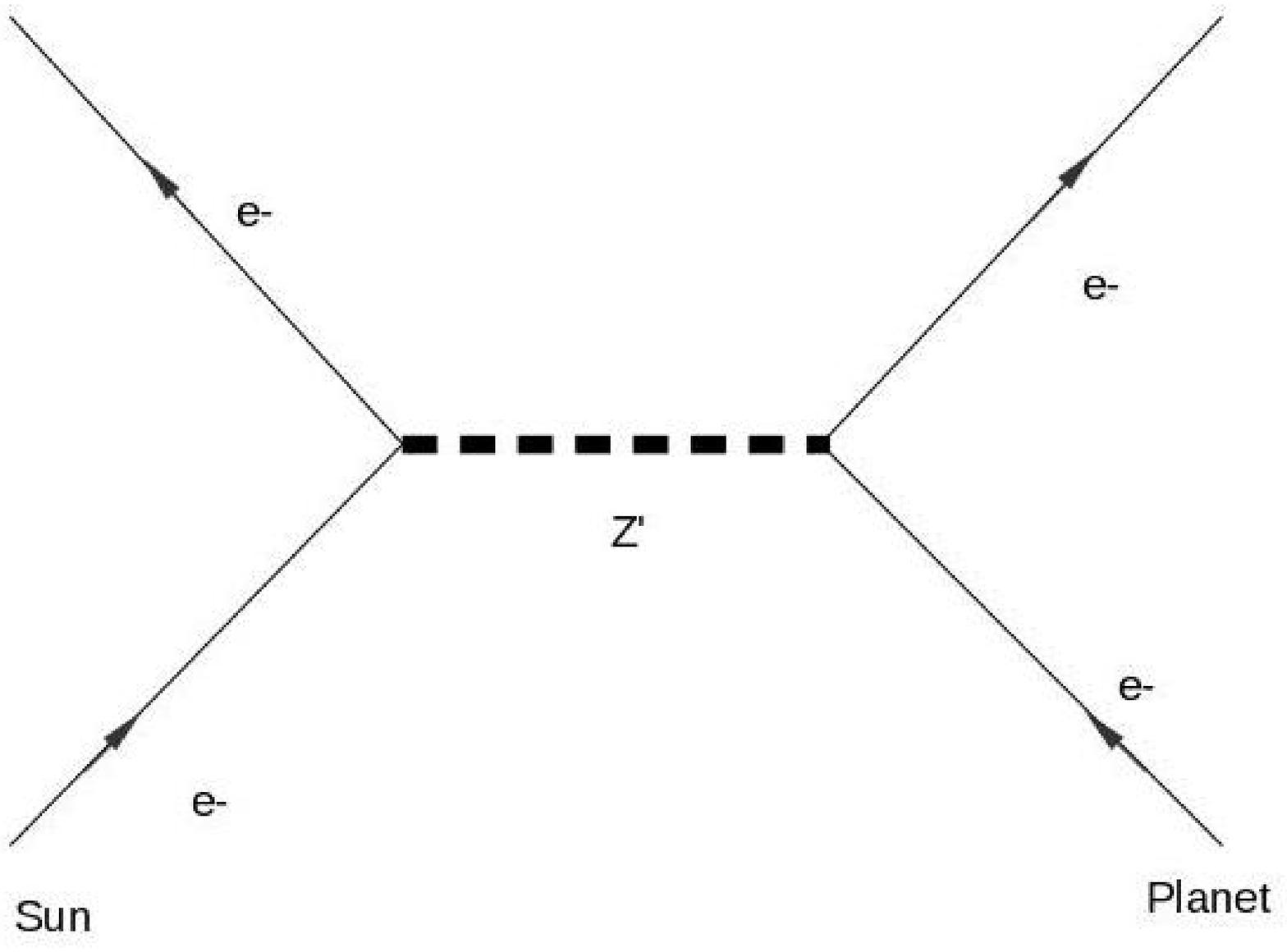}
\caption{Mediation of $L_e-L_{\mu,\tau}$ vector gauge bosons between planet and Sun.}
\label{fig:second}
\end{figure}

In standard model, we can construct three gauge symmetries $L_e-L_\mu$, $L_e-L_\tau$, $L_\mu-L_\tau$ in an anomaly free way and they can be gauged \cite{foo,volkas,foot,dutta}. $L_e-L_\mu$ and $L_e-L_\tau$ \cite{grifols, joshipura, dighe, agarwalla} long range forces can be probed in a neutrino oscillation experiment. $L_\mu-L_\tau$ long range force cannot be probed in neutrino oscillation experiment because Earth and Sun do not contain any muon charge. However, if there is an inevitable $Z-Z^\prime$ mixing, then $L_\mu-L_\tau$ force can be probed \cite{rode}. Recently in \cite{tanmay, toby}, $L_\mu-L_\tau$ long range force was probed from the orbital period decay of neutron star-neutron star and neutron star-white dwarf binary systems since they contain large muon charge. However, as the Sun and the planets contain lots of electrons and the number of electrons is approximately equal to the number of baryons, we can probe $L_e-L_{\mu,\tau}$ long range force from the Solar System. The number of electrons in i'th macroscopic object (Sun or planet) is given by $N_i=M_i/m_n$, where $M_i$ is the mass of the i'th object and $m_n$ is the mass of nucleon which is roughly $1\rm{GeV}$. $L_e-L_{\mu,\tau}$ gauge boson is mediated between the classical electron current sources: Sun and planet as shown in FIG.\ref{fig:second}. This causes a fifth force between the planet and the Sun along with the gravitational force and contributes to the  perihelion shift of the planets. The Yukawa type of potential in such a scenario is $V(r)\simeq\frac{g^2}{4\pi r} e^{-M_{Z^\prime}r}$, where $g$ is the constant of coupling between the electron and the gauge boson and $M_{Z^\prime}$ is the mass of the gauge boson. $M_{Z^\prime}$ is restricted by the distance between the Sun and the planet which gives the strongest bound on gauge boson mass $M_{Z^\prime}<10^{-19}\rm{eV}$. Therefore, the lower bound of the range of this force is given by $\lambda=\frac{1}{M_{Z^\prime}}>10^9\rm{Km}$. $L_e-L_{\mu,\tau}$ long range force can also be probed from MICROSCOPE experiment \cite{f,a,y}. In this mass range the vector gauge boson can also be a candidate for fuzzy dark matter (FDM), although FDM is usually referred to as ultralight scalars \cite{hubarkana,hui}.

The paper is organised as follows. In section II, we give a detail calculation of the perihelion precession of planets due to such fifth force in the background of the Schwarzschild geometry around the Sun. In section III, we obtain constraints on the $L_e-L_{\mu,\tau}$ gauge coupling and the mass of the gauge boson for planets Mercury, Venus, Earth, Mars, Jupiter, and Saturn and we obtain the exclusion plot of $g$ versus $M_{Z^\prime}$ for all the planets mentioned before. In section IV, we summarize our results. We use the natural system of units throughout the paper.  
\section{Perihelion precession of planets due to  long range Yukawa type of potential in the Schwarzschild spacetime background}
The dynamics of a Sun-planet system in presence of a Schwarzschild background and a non gravitational Yukawa type $L_e-L_{\mu,\tau}$ long range force is given by the following action:
\begin{equation}
S=-M_p\int \sqrt{-g_{\mu\nu}\dot{x^\mu}\dot{x^\nu}}d\tau-g\int A_\mu J^\mu d\tau,
\label{eq:action}
\end{equation}
where $``\, \dot{} \,$" (overdot) denotes the derivative with respect to the proper time $\tau$, $g_{\mu\nu}$ is the metric tensor for the background spacetime, $M_p$ is the mass of the planet, $g$ is the coupling constant which couples the classical current  $J^\mu=q\dot{x^\mu}$ of the planet with the $L_e-L_{\mu,\tau}$ gauge field $A_\mu$ due to the Sun, and $q$ is the total charge due to the presence of electrons in the planet. Varying the action Eq.~(\ref{eq:action}), we obtain the equation of motion of the planet as
\begin{equation}
\ddot{x^\alpha}+\Gamma^\alpha_{\mu\nu}\dot{x^\mu}\dot{x^\nu}=\frac{gq}{M_p}g^{\alpha\mu}(\partial_\mu A_\nu-\partial_\nu A_\mu)\dot{x^\nu}.
\label{eq:eom}
\end{equation} 
In Appendix A, we show the detailed calculation of Eq.~(\ref{eq:eom}). For the static case $A_\mu=\{V(r),0,0,0\}$, where $V(r)$ is the potential leading to a long range $L_e-L_{\mu,\tau}$ Yukawa type force. $\Gamma^\alpha_{\mu\nu}$ denotes the Christoffel symbol for the background spacetime. For the Sun-Planet system, the background is a Schwarzschild spacetime outside the Sun and it is described by the line element
\begin{equation}
ds^2=-\Big(1-\frac{2M}{r}\Big)dt^2+\Big(1-\frac{2M}{r}\Big)^{-1}dr^2+r^2 d\theta^2+r^2\sin^2\theta d\phi^2,
\label{eq:b}
\end{equation} 
where $M$ is the mass of the Sun. The Christoffel symbols for the metric  Eq.~(\ref{eq:b}) are given in Appendix B.

Hence, to obtain the solution for temporal part of the Eq.~(\ref{eq:eom}), we write
\begin{equation}
\ddot{t}+\frac{2M}{r^2\Big(1-\frac{2M}{r}\Big)}\dot{r}\dot{t}=\frac{gq}{M_p\Big(1-\frac{2M}{r}\Big)}\frac{dV}{dr}\dot{r}.
\label{eq:kb}
\end{equation}
Integrating Eq.~(\ref{eq:kb}) once, we get
\begin{equation}
\dot{t}=\frac{\Big(E+\frac{gqV}{M_p}\Big)}{\Big(1-\frac{2M}{r}\Big)},
\label{eq:tdot}
\end{equation}
where $E$ is the constant of motion. $E$ is interpreted as the total energy per unit rest mass for a timelike geodesic relative to a static observer at infinity.

Similarly, the $\phi$ part of Eq.~(\ref{eq:eom}) is
\begin{equation}
\ddot{\phi}+\frac{2}{r}\dot{r}\dot{\phi}=0.
\end{equation}
After integration we get
\begin{equation}
\dot{\phi}=\frac{L}{r^2},
\label{eq:phidot}
\end{equation}
where $L$ is the angular momentum of the system per unit mass, which is also a constant of motion. 

The radial part of Eq.~(\ref{eq:eom}) is
\begin{equation}
\ddot{r}-\frac{M\dot{r}^2}{r^2\Big(1-\frac{2M}{r}\Big)}+\frac{M\Big(1-\frac{2M}{r}\Big)}{r^2}\dot{t}^2-r\Big(1-\frac{2M}{r}\Big)\dot{\phi}^2=\frac{gq}{M_p}\Big(1-\frac{2M}{r}\Big)\frac{dV}{dr}\dot{t}.
\label{eq:rdot}
\end{equation}
Using Eqs.~(\ref{eq:tdot}) and (\ref{eq:phidot}) in Eq.~(\ref{eq:rdot}), we obtain
\begin{equation}
\ddot{r}+\frac{M}{r^2\Big(1-\frac{2M}{r}\Big)}\Big(\Big(E+\frac{gqV}{M_p}\Big)^2-\dot{r}^2\Big)-\frac{L^2}{r^3}\Big(1-\frac{2M}{r}\Big)=\frac{gq}{M_p}\Big(E+\frac{gqV}{M_p}\Big)\frac{dV}{dr}.
\label{eq:ti}
\end{equation} 
Again, for a timelike particle $g_{\mu\nu}\dot{x^\mu}\dot{x^\nu}=-1$ and this gives
\begin{equation}
\frac{\Big(E+\frac{gqV}{M_p}\Big)^2-1}{2}=\frac{\dot{r}^2}{2}+\frac{L^2}{2r^2}-\frac{ML^2}{r^3}-\frac{M}{r}.
\label{eq:time}
\end{equation}
Using Eq.~(\ref{eq:time}) in Eq.~(\ref{eq:ti}), we get
\begin{equation}
\ddot{r}+\frac{3ML^2}{r^4}+\frac{M}{r^2}-\frac{L^2}{r^3}=\frac{gq}{M_p}\Big(E+\frac{gqV}{M_p}\Big)\frac{dV}{dr}.
\label{eq:final}
\end{equation} 
We can also obtain Eq.~(\ref{eq:final}) by directly differentiating Eq.~(\ref{eq:time}).

The potential $V(r)$ is generated due to the presence of electrons in the Sun and it is given as $V(r)\simeq \frac{gQ}{4\pi r}e^{-M_{Z^\prime} r}+\mathcal{O}(\frac{M}{R})$, where $R$ is the radius of the Sun. Note that we keep only the Yukawa term in the form of $V(r)$ as we are interested in the leading order contribution only (see Appendix C). Hence, from Eq.~(\ref{eq:time}) we write
\begin{equation}
\frac{E^2-1}{2}=\frac{\dot{r}^2}{2}+\frac{L^2}{2r^2}-\frac{ML^2}{r^3}-\frac{M}{r}-\frac{g^2N_1N_2E}{4\pi M_pr}e^{-M_{Z^\prime} r},
\label{eq:E}
\end{equation}
where we have neglected $\mathcal{O}(g^4)$ term because the coupling is small and its contribution will be negligible. Here $Q=N_1$ is the number of electrons in the Sun and $q=N_2$ is the number of electrons in the planet. For planar motion, $L_x=L_y=0$, and $\theta=\pi/2$. The orbit of the planet is stable when $E<1$.  In the presence of gravitational potential and fifth force $E= E\simeq 1-\frac{M}{2a}+\frac{g^2 Qq}{4\pi M_p}\left(\frac{u_{+}u_{-}^2e^{-M_{Z'}/u_{+}}-u_{+}^2u_{-}e^{-M_{Z'}/u_{-}}}{u_{+}^2-u_{-}^2}\right)$ which is explained in Appendix D.


The first term on the right hand side of Eq.~(\ref{eq:E}) represents the kinetic energy part, the second term is the centrifugal potential part, and the fourth term is the usual Newtonian potential. Due to general relativistic $\frac{ML^2}{r^3}$ term, there is an advancement of perihelion motion of a planet. The last term arises due to exchange of a $U(1)_{L_e-L_{\mu,\tau}}$ gauge bosons between electrons of a planet and the Sun. Here, $M_{Z^\prime}$ is the mass of the gauge boson. $M_{Z^\prime}$ is constrained from the range of the potential which is basically the distance between the planet and the Sun. Using $\dot{r}=\frac{L}{r^2}\frac{dr}{d\phi}$, we write Eq.~(\ref{eq:E}) as
\begin{equation}
\Big[\frac{d}{d\phi}\Big(\frac{1}{r}\Big)\Big]^2+\frac{1}{r^2}=\frac{E^2-1}{L^2}+\frac{2M}{r^3}+\frac{2M}{L^2 r}+\frac{g^2N_1N_2E}{2\pi L^2r M_p}e^{-M_{Z^\prime} r}.
\label{eq:e}
\end{equation}
Applying $\frac{d}{d\phi}$ on both sides and using the reciprocal coordinate $u=\frac{1}{r}$ we obtain from Eq.~(\ref{eq:e})
\begin{equation}
\frac{d^2 u}{d\phi^2}+u=\frac{M}{L^2}+3Mu^2+\frac{g^2N_1N_2}{4\pi L^2M_p}e^{-\frac{M_{Z^\prime}}{u}}+\frac{g^2N_1N_2E M_{Z^\prime}}{4\pi L^2 M_p u}e^{-\frac{M_{Z^\prime}}{u}}.
\label{eq:f}
\end{equation}
 As $E$ appears as a multiplication factor in Eq.~(\ref{eq:f}), we take $E\approx 1$ as other terms are very small. Hence, expanding Eq.~(\ref{eq:f}) upto the leading order of $M_{Z^\prime}$, we get
\begin{equation}
\frac{d^2 u}{d\phi^2}+u=\frac{M}{L^2}+3Mu^2+\frac{g^2N_1N_2}{4\pi L^2M_p}-\frac{g^2N_1N_2M^2_{Z^\prime}}{8\pi L^2M_pu^2},
\label{eq:ff}
\end{equation}

where for non circular orbit $\frac{d}{d\phi}\Big(\frac{1}{r}\Big)\neq 0$. The first term on the right hand side of Eq.~(\ref{eq:ff}) is the usual term which comes in Newton's theory. The second term is the general relativistic term which is a perturbation of Newton's theory. The last two terms arise due to the presence of long range Yukawa type potential in the theory.

We write Eq.~(\ref{eq:ff}) as
\begin{equation}
\frac{d^2 u}{d\phi^2}+u=\frac{M^\prime}{L^2}+3Mu^2-\frac{g^2N_1N_2M^2_{Z^\prime}}{8\pi L^2M_pu^2},
\end{equation}
where $M^\prime=M+g^2N_1N_2/4\pi M_p$.

We assume that $u=u_0(\phi)+\Delta u(\phi)$, where, $u_0(\phi)$ is the solution of Newton's theory with the effective mass $M^\prime$ and $\Delta u(\phi)$ is the solution due to general relativistic correction and Yukawa potential. Thus we write
\begin{equation}
\frac{d^2 u_0}{d\phi^2}+u_0=\frac{M^\prime}{L^2}.
\label{eq:g}
\end{equation} 
The solution of Eq.~(\ref{eq:g}) is
\begin{equation}
u_0=\frac{M^\prime}{L^2}(1+e\cos\phi),
\label{eq:one}
\end{equation}
where $e$ is the eccentricity of the planetary orbit. The equation of motion for $\Delta u(\phi)$ is
\begin{equation}
\frac{d^2 \Delta u}{d\phi^2}+\Delta u=\frac{3MM^{\prime^2}}{L^4}(1+e^2\cos^2\phi+2e\cos\phi)-\frac{g^2N_1N_2M^2_{Z^\prime}L^4}{8\pi L^2M_pM^{\prime^2}(1+e^2\cos^2\phi+2e\cos\phi)}.
\label{eq:h}
\end{equation}
The solution of Eq.~(\ref{eq:h}) is
\begin{equation}
\begin{split}
\Delta u=\frac{3MM^{\prime^2}}{L^4}\Big[1+\frac{e^2}{2}-\frac{e^2}{6}\cos2\phi+e\phi\sin\phi\Big]-\frac{g^2N_1N_2M^2_{Z^\prime} L^4}{8\pi L^2M_pM^{\prime^2}}\Big[
-\frac{\cos\phi}{e(1+e\cos\phi)}+\\
\frac{\sin^2\phi}{(1-e^2)(1+e\cos\phi)}-\frac{e}{(1-e^2)^{3/2}}\sin\phi\cos^{-1}\Big(\frac{e+\cos\phi}{1+e \cos\phi}\Big)\Big].
\end{split}
\label{eq:i}
\end{equation}
When $\Delta u$ increases linearly with $\phi$, it contributes to the perihelion precession of planets. Therefore, we identify only the related terms in Eq.~(\ref{eq:i}), neglect all other terms, and rewrite $\Delta u$ as
\begin{equation}
\Delta u=\frac{3MM^{\prime^2}}{L^4}e\phi\sin\phi+\frac{g^2N_1N_2M^2_{Z^\prime} L^2}{8\pi M_p M^{\prime^2}}\frac{e}{(1-e^2)(1+e)}\phi\sin\phi,
\label{eq:j}
\end{equation}
where we used $\cos^{-1}\Big(\frac{e+\cos\phi}{1+e \cos\phi}\Big)\simeq \frac{\sqrt{1-e^2}}{1+e}\phi +\mathcal{O}(\phi^2)$.

Using Eqs.~(\ref{eq:one}) and (\ref{eq:j}), we get the total solution as
\begin{equation}
u=\frac{M^\prime}{L^2}(1+e\cos\phi)+\frac{3MM^{\prime^2}}{L^4}e\phi\sin\phi+\frac{g^2N_1N_2 M^2_{Z^\prime} L^2}{8\pi M_p M^{\prime^2}}\frac{e}{(1-e^2)(1+e)}\phi\sin\phi,
\label{eq:k}
\end{equation}
or,
\begin{equation}
u=\frac{M^\prime}{L^2}[1+e\cos\phi(1-\alpha)],
\end{equation}
where,
\begin{equation}
\alpha=\frac{3MM^\prime}{L^2}+\frac{g^2N_1N_2M^2_{Z^\prime} L^4}{8\pi M_p M^{\prime^3}}\frac{1}{(1-e^2)(1+e)}.
\label{eq:km}
\end{equation}
Under $\phi\rightarrow \phi+2\pi$, $u$ is not same. Hence, the planet does not follow the previous orbit. So the motion of the planet is not periodic. The change in azimuthal angle after one precession is
\begin{equation}
\Delta \phi=\frac{2\pi}{1-\alpha}-2\pi\approx 2\pi\alpha.
\label{eq:l}
\end{equation}
The semi major axis and the orbital angular momentum are related by $a=\frac{L^2}{M^\prime(1-e^2)}$. Using this expression in Eq.~(\ref{eq:l}) we get
\begin{equation}
\Delta \phi=\frac{6\pi M}{a(1-e^2)}+\frac{g^2N_1N_2M^2_{Z^\prime} a^2(1-e^2)}{4M_p M^\prime(1+e)}.
\label{eq:m}
\end{equation}
In natural system of units Eq.~(\ref{eq:m}) is 
\begin{equation}
\Delta \phi=\frac{6\pi GM}{a(1-e^2)}+\frac{g^2N_1N_2M^2_{Z^\prime} a^2(1-e^2)}{4M_p (GM+\frac{g^2N_1N_2}{4\pi M_p})(1+e)}.
\label{eq:n}
\end{equation}
The energy due to gravity is much larger than the energy due to long range Yukawa type force. The last term of Eq.~(\ref{eq:n}) indicates that long range force, which arises due to $U(1)_{L_e-L_{\mu,\tau}}$ gauge boson exchange between the electrons of composite objects, contributes to the perihelion advance of planets within the permissible limit. 
\section{Constraints on $U(1)_{L_e-L_{\mu,\tau}}$ gauge coupling for planets in Solar system }  
The contribution of the gauge boson must be within the excess of perihelion advance from the GR prediction, i.e; $(\Delta \phi)_{obs}-(\Delta\phi)_{GR} \geq (\Delta\phi)_{L_e-L_{\mu,\tau}}$. The first term in the right hand side of Eq.~(\ref{eq:n}) is $(\Delta\phi)_{GR}$ and the second term is $(\Delta\phi)_{L_e-L_{\mu,\tau}}$. Putting the observed and GR values for $(\Delta\phi)$, we can constrain the $U(1)_{L_e-L_{\mu,\tau}}$ gauge coupling constants for all the planets in our Solar System. For Mercury planet, we write
\begin{equation}
\frac{g^2N_1N_2M^2_{Z^\prime} a^2(1-e^2)}{4M_p (GM+\frac{g^2N_1N_2}{4\pi M_p})(1+e)}\Big(\frac{{\rm century}}{T}\Big)<3.0\times 10^{-3} {\rm arcsecond/century},
\end{equation}
where $3\times 10^{-3}$ arcsecond/century is the uncertainty in the perihelion advancement from its GR prediction and put upper bound on the gauge coupling. $T=88$ days is the orbital time period of Mercury. Similarly, we can put upper bounds on $g$ for other planets.
In this section, we constrain the $U(1)_{L_e-L_{\mu,\tau}}$ gauge coupling from the observed perihelion advancement of the planets in the Solar System. We consider six planets: Mercury, Venus, Earth, Mars, Jupiter, and Saturn. Here, we take the mass of the Sun as $M=10^{57}$GeV. Using  Eqs.~(\ref{eq:n}),  we put an upper bound on $g$ from the uncertainty of their perihelion advance. In TABLE \ref{tableI}, we obtain the upper bound on masses of the gauge bosons which are mediated between the Sun and the planets and, in TABLE \ref{tableII}, we show the constraints on the gauge coupling constants from the uncertainties \cite{cwill,pitjeva} of perihelion advance.

\begin{table}[h]
\caption{\label{tableI}Summary of the masses, eccentricities \cite{url} of the orbits, perihelion distances from the Sun and upper bounds on gauge boson mass $M_{Z^\prime}$ which are mediated between the planets and Sun in our Solar System.}
\centering
\begin{tabular}{ |l|c|c|c|c|  }
 
 \hline
Planet \hspace{0.01cm} & Mass $M_p$(GeV)\hspace{0.01cm}& Eccentricity (e)\hspace{0.01cm} & Perihelion distance a (AU) \hspace{0.01cm} &Mass of gauge boson $M_{Z^\prime}$(eV)\\
 \hline
Mercury&$1.84\times 10^{50}$  & $0.206$  & $0.31$&$\leq4.26\times 10^{-18}$ \\
Venus&$2.73\times 10^{51}$ & $0.007$  &$0.72$&$\leq1.83\times 10^{-18}$\\
Earth &$3.35\times 10^{51}$ & $0.017$  & $0.98$&$\leq1.35\times 10^{-18}$ \\
Mars &$3.59\times 10^{50}$&  $0.093$  &$1.38$&$\leq9.56\times 10^{-19}$\\
Jupiter &$1.07\times 10^{54}$& $0.048$  &$4.95$&$\leq2.67\times10^{-19}$\\
Saturn &$3.19\times 10^{53}$& $0.056 $  &$9.02$&$\leq1.46\times 10^{-19}$\\
 \hline
\end{tabular}
\end{table}

\begin{figure}[!htbp]
\centering
\includegraphics[width=5.0in,angle=360]{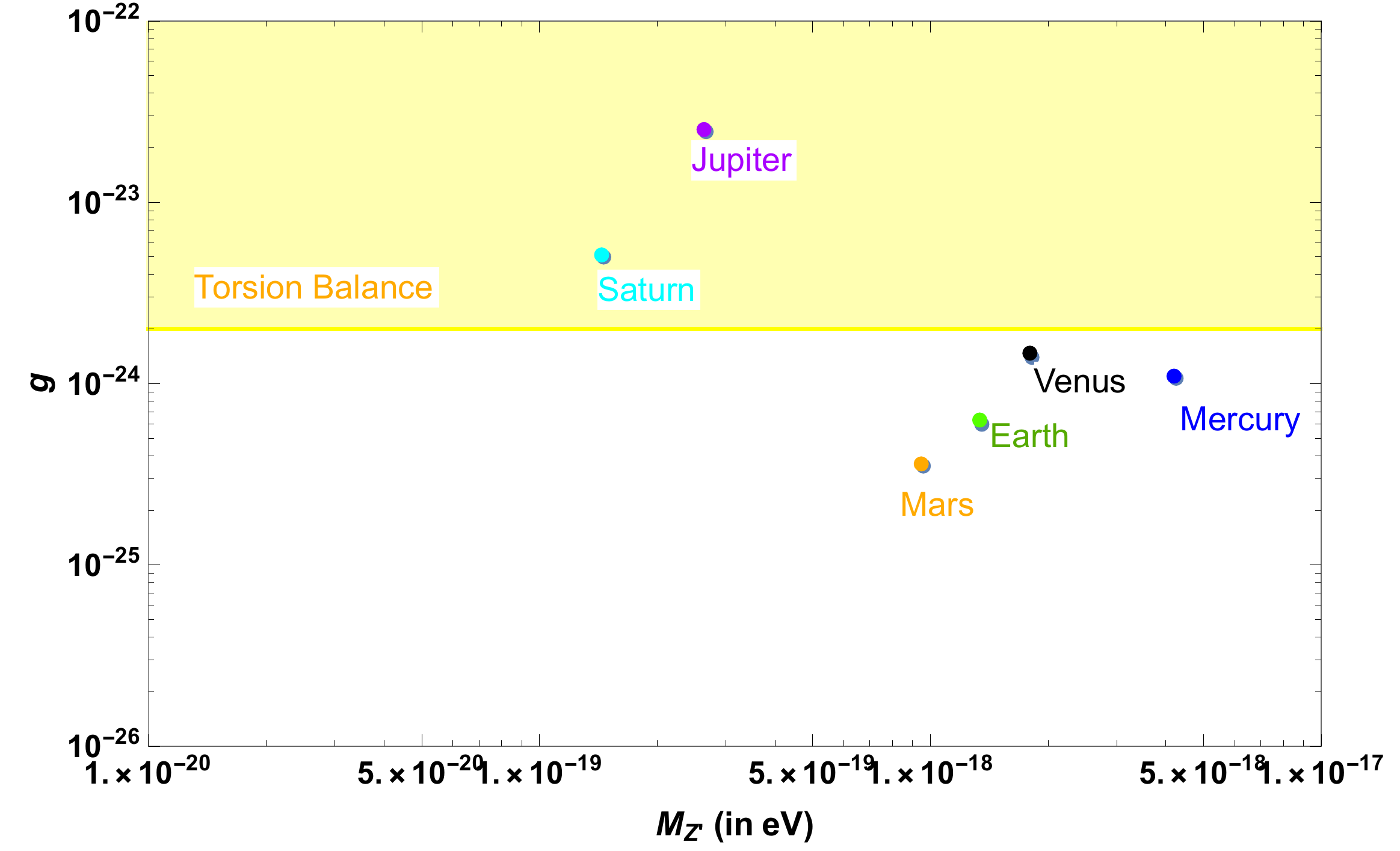}
\caption{Values of the gauge coupling of each planets corresponding to the Sun-planet distance obtained from TABLE\ref{tableII}. Violet dot is for Jupiter planet, blue dot is for Mercury planet, black dot is for Venus, cyan dot is for Saturn, green dot is for Earth and yellow dot is for Mars. The yellow shaded region is excluded from the torsion balance experiments}
\label{fig:planet1}
\end{figure}

\begin{figure}[!htbp]
\centering
\includegraphics[width=5.0in,angle=360]{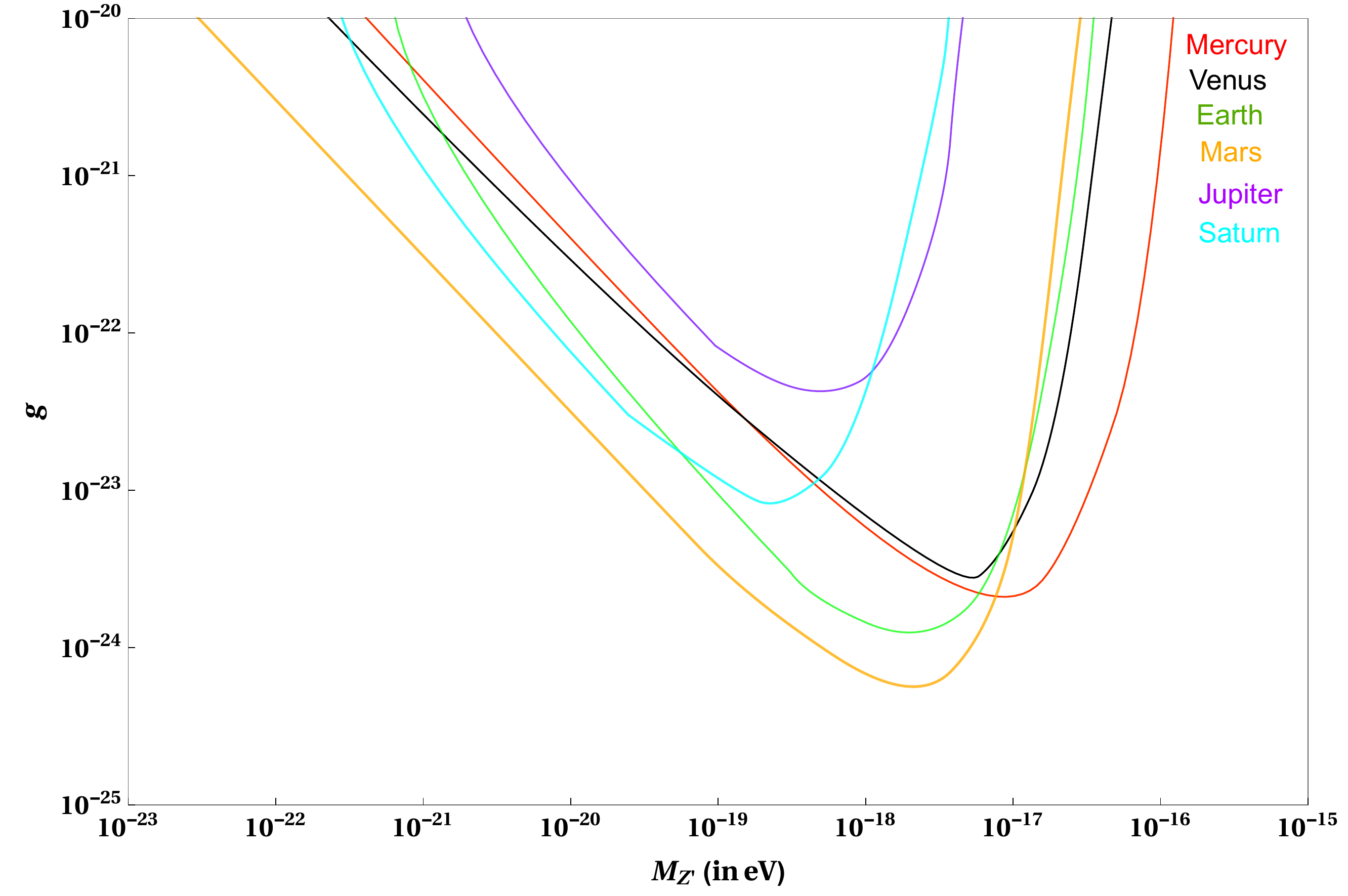}
\caption{Plot of coupling constant $g$ vs the mass of the gauge bosons $M^\prime_Z$ for all the planets. Violet line is for Jupiter planet, red line is for Mercury planet, black line is for Venus, cyan line is for Saturn, green line is for Earth and yellow line is for Mars.}
\label{fig:planet}
\end{figure}

\begin{table}[h]
\caption{\label{tableII} Summary of the uncertainties in the perihelion advance in arcseconds per century and upper bounds on gauge boson-electron coupling $g$ for the values of $M_{Z^\prime}$ discussed in TABLE\ref{tableI}  for planets in our Solar System.}
\centering
\begin{tabular}{ |l|c|c|  }
 
 \hline
Planet \hspace{0.1cm} & Uncertainty in perihelion advance (as/cy)\hspace{0.1cm} & $g$ from perihelion advance\\
 \hline
Mercury  & $3.0\times 10^{-3}$  & $\leq1.055\times 10^{-24}$\\
Venus & $1.6\times 10^{-3}$  &$\leq1.377\times 10^{-24}$\\
Earth & $1.9\times 10^{-4}$  & $\leq6.021\times 10^{-25}$ \\
Mars &  $3.7\times 10^{-5}$  &$\leq3.506\times 10^{-25}$\\
Jupiter & $2.8\times 10^{-2} $  &$\leq2.477\times 10^{-23}$\\
Saturn & $4.7\times 10^{-4}$  &$\leq5.040\times 10^{-24}$\\
 \hline
\end{tabular}
\end{table}
We can write from the fifth force constraint
\begin{equation}
\frac{g^2 N_1 N_2}{4\pi G M M_p}<1.
\label{eq:yt}
\end{equation}
This gives the upper bound on $g$ as $g<3.54\times 10^{-19}$ for all the planets. In FIG.\ref{fig:planet1} we show the values of gauge coupling of the planets corresponding to the planet-Sun distance.
For $U(1)_{L_e-L_{\mu,\tau}}$ vector gauge bosons exchange between the planet and the Sun, the mass of the gauge boson is $M_{Z^\prime}\leq \mathcal{O}(10^{-19})eV$. In FIG.\ref{fig:planet}, we obtain the exclusion plots of gauge boson electron coupling for the six planets by numerically solving Eqs.~(\ref{eq:f}). There is an extra multiplicative factor $\exp[{\frac{-M^\prime_Z L^2}{M^\prime}}]$ in the expression of $\alpha$ if we solve Eqs.~(\ref{eq:f}) numerically in order to incorporate the exponential suppresion due to higher values of $M_{Z^{\prime}}$. 
The regions above the coloured lines corresponding to every planets are excluded. Eqs.~(\ref{eq:n}) suggests that the perihelion shift due to the mediation of $L_e-L_{\mu,\tau}$ gauge bosons is proportional to the square of the semi major axis. This is completely opposite from the standard GR result where the perihelion shift is inversely proportional to $a$ for small $M_{Z^{\prime}}$. However, for higher values of $M_{Z^{\prime}}$, the exponential suppression starts dominating. So the contribution of the gauge boson mediation for perihelion shift is larger for outer planets. However it also depends on the available uncertainties for perihelion precession of the planets and other parameters like orbital time period and eccentricity. From TABLE \ref{tableII}, we obtain the stronger bound on the gauge boson coupling is $g\leq\mathcal{O}(10^{-25})$. From FIG.\ref{fig:planet} it is clear that the Mars gives the strongest bound among all the planets considered. As we go to the lower mass region, the exponential term in the potential will become less effective and the Yukawa potential effectively becomes Coulomb potential at $M_{Z'}\rightarrow 0$. Thus it will be degenerate with $1/r^2$-Newtonian force and will not contribute to the perihelion precession of planets at all. 
So as we go to the lower mass $(<10^{-19}eV)$ region, we get weaker bound on $g$. On the other hand, for higher mass region $(>10^{-19}eV)$ the long range force theory breaks down and, thus we can not go arbitrarily for higher masses. 
\section{Discussions}
Since the Sun and the planets contain a significant number of electrons, long range Yukawa type fifth force can be mediated between the electrons of Sun and planet in a gauged $L_e-L_{\mu,\tau}$ scenario. Also there can be the dipole radiation of the gauge bosson for the planeraty orbits. Following our previous work \cite{tanmay} on compact binary systems in a gauged  $L_{\mu}-L_{\tau}$ scenario, the energy loss due to dipole radiation is proportional to the fourth power of the orbital frequency. For planet-Sun binary system, the orbital frequency is smaller than the orbital frequency of the compact binary systems. Hence, the contribution due to dipole radiation for the planetary systems is smaller and its effect will be neglected for planetary motion.

This ultralight vector gauge bosons mediated between the Sun and the planets can contribute to the perihelion shift in addition to the GR prediction. From the perihelion shift calculation in presence of a long range Yukawa type potential, we obtain an upper bound on the gauge coupling $g\leq\mathcal{O}(10^{-25})$ in a gauged $L_e-L_{\mu,\tau}$ scenario. The mass of the gauge bosons is constrained by the distance between the Sun and the planet which gives $M_{Z^\prime}\leq{\mathcal{O}(10^{-19})}\rm {eV}$. The electron-gauge boson coupling obtained from perihelion shift measurement is six order of magnitude more stringent than our fifth force constraint Eq.~(\ref{eq:yt}). From Eq.~(\ref{eq:n}) we conclude that, while the precession of perihelion due to GR is largely contributed by the planets close to Sun, the contribution of vector gauge bosons in perihelion precession is dominated by the outer planets. 

The bound on coupling $g$ that we have obtained is not only as good as the torsion balance \cite{torsion} or the neutrino oscillation experiment \cite{joshipura}, but also our results possess additional importance for the following reasons:
\begin{itemize}
\item[$(a)$]
Our analysis of the perihelion precession is sensitive to the magnitude of the potential and the nature of the potential, i.e. the deviation from the inverse square law.

\item[$(b)$]In our analysis, we are probing larger distance (upto the planet Saturn) compare to the earth Sun distance.

\item[$(c)$] Since the perihelion shift depends on the value of uncertainty in GR prediction, the future BepiColombo mission \cite{will} can give more accurate result and the bound on coupling will become even more stronger.
\end{itemize} 
Moreover, we emphasize the novel physics behind the work which suggests that we can study the gauge boson electron coupling in a gauged $L_e - L_{\mu,\tau}$ scenario by planetary observations and we can constrain the arising long range force from perihelion precession of planets. These gauge bosons ($M_{Z^\prime}\leq{10^{-19}}\rm {eV}$) can be a possible candidate of fuzzy dark matter and can be probed from precession measurement of planetary orbits.


\section*{Acknowledgments}

SJ was supported by the Swiss Government Excellence Scholarship 2019 (Postdoctoral) for foreign researchers offered via the Federal Commission for Scholarships (FCS) for Foreign Students.
  
\appendix
\section{Equation of motion of a planet in presence of a Schwarzschild background and a non gravitational Yukawa type of potential}
The action which describes the motion of a planet in Schwarzschild background and a non gravitational long range Yukawa type of potential is given by Eq.~(\ref{eq:action}).

Suppose $S_1=M_p\int \sqrt{-g_{\mu\nu}\dot{x^\mu}\dot{x^\nu}}d\tau$. For this action, the Lagrangian is
\begin{equation}
\mathcal{L}=M_p\sqrt{g_{\mu\nu}\frac{dx^\mu}{d\tau}\frac{dx^\nu}{d\tau}}.
\end{equation}
Hence, the equation of motion is
\begin{equation}
\frac{d}{d\tau}\Big(\frac{\partial\mathcal{L}}{\partial \big(\frac{\partial x^\sigma}{\partial \tau}\big)}\Big)-\frac{\partial \mathcal{L}}{\partial x^\sigma}=0,
\end{equation}
or,
\begin{equation}
\frac{1}{\mathcal{L}}\frac{d\mathcal{L}}{d\tau}g_{\mu\sigma}\frac{dx^\mu}{d\tau}=g_{\mu\sigma}\frac{d^2x^\mu}{d\tau^2}+\partial_\alpha g_{\mu\sigma}\frac{dx^\alpha}{d\tau}\frac{dx^\mu}{d\tau}-\frac{1}{2}\partial_\sigma g_{\mu\nu}\frac{dx^\mu}{d\tau}\frac{dx^\nu}{d\tau}.
\end{equation}
Multiplying $g^{\rho\sigma}$ we have,
\begin{equation}
\frac{d^2x^\rho}{d\tau^2}+g^{\rho\sigma}\partial_\nu g_{\mu\sigma}\frac{dx^\nu}{d\tau}\frac{dx^\mu}{d\tau}-g^{\rho\sigma}\frac{1}{2}\partial_\sigma g_{\mu\nu}\frac{dx^\mu}{d\tau}\frac{dx^\nu}{d\tau}=\frac{1}{\mathcal{L}}\frac{d\mathcal{L}}{d\tau}\frac{dx^\rho}{d\tau},
\label{eq:a1}
\end{equation}
or,
\begin{equation}
\frac{d^2x^\rho}{d\tau^2}+\frac{1}{2}g^{\rho\sigma}(\partial_\nu g_{\mu\sigma}+\partial_\mu g_{\nu\sigma}-\partial_\sigma g_{\mu\nu})\frac{dx^\mu}{d\tau}\frac{dx^\nu}{d\tau}=\frac{1}{\mathcal{L}}\Big(\frac{d\mathcal{L}}{d\tau}\Big)\frac{dx^\rho}{d\tau},
\end{equation}
or,
\begin{equation}
\frac{d^2x^\rho}{d\tau^2}+\Gamma^\rho_{\mu\nu}\frac{dx^\mu}{d\tau}\frac{dx^\nu}{d\tau}=\frac{1}{\mathcal{L}}\frac{d\mathcal{L}}{d\tau}\frac{dx^\rho}{d\tau},
\end{equation}
where, $\Gamma^\rho_{\mu\nu}=\frac{1}{2}g^{\rho\sigma}(\partial_\nu g_{\mu\sigma}+\partial_\mu g_{\nu\sigma}-\partial _\sigma g_{\mu\nu})$ is called the Christoffel symbol. We can choose $\tau$ in such a way that $\frac{d\mathcal{L}}{d\tau}=0$. This is called affine parametrization. So,
\begin{equation}
\frac{d^2x^\rho}{d\tau^2}+\Gamma^\rho_{\mu\nu}\frac{dx^\mu}{d\tau}\frac{dx^\nu}{d\tau}=0.
\label{eq:a2}
\end{equation}
Suppose $S_2=gq\int A_\mu \frac{dx^\mu}{d\tau}d\tau=gq\int A_\mu dx^\mu$. Hence,
\begin{equation}
\delta S_2=gq\int \delta A_\mu dx^\mu+gq\int A_\mu \delta(dx^\mu),
\end{equation}
or,
\begin{equation}
\delta S_2=gq\int \frac{\partial A_\mu}{\partial x^\nu}\delta x^\nu dx^\mu+gq\int A_\mu d(\delta x^\mu).
\end{equation}
Using integration by parts and using the fact that the total derivative term will not contribute to the integration, we can write
\begin{equation}
\delta S_2=gq\int\frac{\partial A_\mu}{\partial x^\nu}\delta x^\nu dx^\mu- gq\int dA_\mu \delta x^\mu.
\end{equation}
or,
\begin{equation}
\delta S_2=gq\int\frac{\partial A_\mu}{\partial x^\nu}\delta x^\nu dx^\mu- gq \int \frac{\partial A_\mu}{\partial x^\nu}dx^\nu\delta x^\mu. 
\end{equation}
Since $\mu$ and $\nu$ are dummy indices, we interchange  $\mu$ and $\nu$ in the first term. Hence, we can write
\begin{equation}
\delta S_2=gq\int (\partial _\mu A_\nu-\partial _\nu A_\mu)dx^\nu\delta x^\mu=gq\int (\partial _\mu A_\nu-\partial _\nu A_\mu)\frac{dx^\nu}{d\tau}\delta x^\mu d\tau.
\label{eq:a3}
\end{equation}
Imposing the fact $\delta S_1+\delta S_2=0$ and using Eq.~(\ref{eq:a1}), Eq.~(\ref{eq:a2}) and Eq.~(\ref{eq:a3}) we can write 
\begin{equation}
\ddot{x^\rho}+\Gamma^\rho_{\mu\nu}\dot{x^\mu}\dot{x^\nu}=\frac{gq}{M_p}g^{\rho\mu}(\partial_\mu A_\nu-\partial_\nu A_\mu)\dot{x^\nu},
\end{equation} 
which matches with Eq.~(\ref{eq:eom}).
\section{Christoffel symbols for the Schwarzschiild metric}
The christoffel symbols for the Schwarzschiild metric defined in Eq.~(\ref{eq:b}) are
\begin{equation}
\begin{split}
\Gamma^t_{rt}=\frac{M}{r^2\Big(1-\frac{2M}{r}\Big)},\hspace{0.2cm} \Gamma^r_{tt}=\frac{M}{r^2}\Big(1-\frac{2M}{r}\Big),\hspace{0.2cm} \Gamma^r_{rr}=-\frac{M}{r^2\Big(1-\frac{2M}{r}\Big)} ,\hspace{0.2cm} \Gamma^r_{\theta\theta}=-r\Big(1-\frac{2M}{r}\Big)\\
\Gamma^r_{\phi\phi}=-r\sin^2\theta\Big(1-\frac{2M}{r}\Big),\hspace{0.2cm} \Gamma^\theta_{r\theta}=\frac{1}{r},\hspace{0.2cm} \Gamma^\theta_{\phi\phi}=-\sin\theta\cos\theta,\hspace{0.2cm} \Gamma^\phi_{\phi r}=\frac{1}{r}, \hspace{0.2cm} \Gamma^\phi_{\theta\phi}=\cot\theta.
\end{split}
\end{equation}
\section{Equation of motion for the vector field $A_\mu$}
The vector field $A_\mu$ satisfies the Klein-Gordon equation 
\begin{equation}
\Box A_\mu=M^2_{Z^\prime}A_\mu.
\label{eq:p1}
\end{equation}
Now, for the static case, $A_\mu=\{V(r),0,0,0\}$. Hence,
\begin{equation}
\Box V(r)=M^2_{Z^\prime}V(r).
\label{eq:p2}
\end{equation}
In the background of the Schwarzschild spacetime, Eq.~(\ref{eq:p2}) becomes
\begin{equation}
\Big(1-\frac{2M}{r}\Big)\frac{d^2V}{dr^2}+\frac{2}{r}\Big(1-\frac{M}{r}\Big)\frac{dV}{dr}=M^2_{Z^\prime}V(r).
\label{eq:p3}
\end{equation}
So, in the Schwarzschild background, $V(r)$ will not satisfy the Klein-Gordon equation. So we expand $V(r)$ in a perturbation series where the perturbation parameter is $\frac{M}{R}$, and the leading order term is the Yukawa term. Let,
\begin{equation}
V(r)=V_0(r)+\frac{M}{R}V_1(r)+\mathcal{O}\Big(\frac{M}{R}\Big)^2,
\label{eq:p4}
\end{equation}
where
\begin{equation}
V_0(r)=c\frac{e^{-M^\prime_{Z}r}}{r},\hspace{1cm} c=\frac{g^2N_1N_2}{4\pi},
\label{eq:p5}
\end{equation}
such that
\begin{equation}
\frac{d^2V_0}{dr^2}+\frac{2}{r}\frac{dV_0}{dr}=M^2_{Z^\prime}V_0.
\label{eq:p6}
\end{equation}
Inserting Eq.~(\ref{eq:p4}) in Eq.~(\ref{eq:p3}), we get the equation for $V_1(r)$
\begin{equation}
\frac{1}{R}\frac{d^2V_1}{dr^2}+\frac{2}{rR}\frac{dV_1}{dr}=\frac{M^2_{Z^\prime}V_1}{R}+\frac{2}{r}\frac{d^2V_0}{dr^2}+\frac{2}{r^2}\frac{dV_0}{dr}.
\label{eq:p7}
\end{equation}
Let,
\begin{equation}
V_1(r)=\chi(r)\frac{e^{-M^\prime_Zr}}{r}.
\label{eq:p8}
\end{equation}
Now, Eq.~(\ref{eq:p7}) becomes
\begin{equation}
\frac{1}{R}\frac{d^2\chi}{dr^2}-\frac{1}{R}2M^\prime_Z\frac{d\chi}{dr}=2c\Big(\frac{M^2_{Z^\prime}}{r}+\frac{1}{r^3}+\frac{M^\prime_Z}{r^2}\Big).
\label{eq:p9}
\end{equation}
Integrating Eq.~(\ref{eq:p9}) once we get
\begin{equation}
\frac{d\chi}{dr}-2M^\prime_Z \chi=2cR\Big[M^2_{Z^\prime}\ln(M^\prime_Zr)-\frac{1}{2r^2}-\frac{M^\prime_Z}{r}\Big]+k_1R,
\label{eq:p10}
\end{equation}
where $k_1$ is the integration constant. Eq.~(\ref{eq:p10}) can be written as
\begin{equation}
\frac{d}{dr}\Big(e^{-2M^\prime_Zr}\chi\Big)=2cRe^{-2M^\prime_Zr}\Big[M^2_{Z^\prime}\ln(M^\prime_Zr)-\frac{1}{2r^2}-\frac{M^\prime_Z}{r}\Big]+k_1Re^{-2M^\prime_Zr}.
\label{eq:p11}
\end{equation} 
From Eq.~(\ref{eq:p11}), we can write
\begin{equation}
e^{-2M^\prime_Zr}\chi(r)=2cR\Big[M^2_{Z^\prime}\int^r_\infty e^{-2M^\prime_Zx}\ln(M^\prime_Zx)dx-\int^r_\infty \frac{e^{-2M^\prime_Zx}}{2x^2}dx-\int^r_\infty\frac{M^\prime_Ze^{-2M^\prime_Zx}}{x}dx\Big]-\frac{k_1R}{2M^\prime_Z}e^{-2M^\prime_Zr}+k_2,
\label{eq:p12}
\end{equation}
where $k_2$ is an integration constant. Doing integration by parts, Eq.~(\ref{eq:p12}) becomes
\begin{equation}
\chi(r)=cR\Big[-M^\prime_Z\ln(M^\prime_Zr)+\frac{1}{r}+M^\prime_Ze^{2M^\prime_Zr}E_i(-2M^\prime_Zr)\Big]-\frac{k_1R}{2M^\prime_Z}+k_2e^{2M^\prime_Zr},
\label{eq:p13}
\end{equation}
where $E_i(x)$ is a special function called the exponential integral function which is defined as
\begin{equation}
E_i(x)=-\int^\infty_{-x} \frac{e^{-t}}{t}dt.
\label{eq:p14}
\end{equation}
We chose $k_2=0$ as $e^{2M^\prime_Zr}$ diverges. We also chose $k_1=0$ as we are looking for particular integral. Hence, from Eq.~(\ref{eq:p13}) we get
\begin{equation}
V_1(r)=\frac{cRe^{-M^\prime_Zr}}{r}\Big[\frac{1}{r}-M^\prime_Z\ln(M^\prime_Zr)+M^\prime_Ze^{2M^\prime_Zr}E_i(-2M^\prime_Zr)\Big].
\label{eq:p15}
\end{equation}
So the total solution of the potential is
\begin{equation}
V(r)=\frac{ce^{-M^\prime_Zr}}{r}\Big[1+\frac{M}{r}\{1-M^\prime_Zr\ln(M^\prime_Zr)+M^\prime_Zre^{2M^\prime_Zr}E_i(-2M^\prime_Zr)\}\Big]+\mathcal{O}\Big(\frac{M^2}{R^2}\Big).
\end{equation}
We take the leading order term which is the Yukawa term in our calculation. The higher order terms are comparatively small. 
\section{Total energy of the binary system due to gravity and long range Yukawa type potential}
For Newtonian gravity, we can write 
\begin{equation}
\frac{E^2-1}{L^2}=-\frac{1}{a^2(1-e^2)}, \hspace{2cm} \frac{2M}{L^2}=\frac{2}{a(1-e^2)}. 
\end{equation}
Dividing the above two expression, we obtain
\begin{equation}
\frac{E^2-1}{M}=-\frac{1}{a},
\end{equation}
or,
\begin{equation}
E\simeq\sqrt{1-\frac{M}{a}}\approx 1-\frac{M}{2a}.
\end{equation}
In presence of long range Yukawa potential, we obtain $E$ from the condition $\frac{du}{d\phi}=0$ at $u=u_{+}=1/{a(1+e)}$ (aphelion) and $u=u_{-}=1/{a(1-e)}$ (perihelion), 
\begin{equation}
E\simeq 1-\frac{M}{2a}+\frac{g^2 Qq}{4\pi M_p}\left(\frac{u_{+}u_{-}^2e^{-M_{Z'}/u_{+}}-u_{+}^2u_{-}e^{-M_{Z'}/u_{-}}}{u_{+}^2-u_{-}^2}\right)
\end{equation}
where $1$ in the right hand side is the rest energy per unit mass in the Minkowski background. The second term is $\approx 10^{-8}$ and the third Yukawa term is smaller than the Newtonian term.

\end{document}